\documentclass[twocolumn,english]{revtex4-2}
\usepackage[T1]{fontenc}
\usepackage[latin9]{inputenc}
\usepackage{amsmath}
\usepackage{amsfonts}
\usepackage{graphicx}
\usepackage{babel}
\begin{document}
\title{Exact spin form factors and correlations at a massive Kramers--Wannier interface}
\author{Zoltan Bajnok$\,^{a,b}$, Yizhuang Liu$
\,^{b}$}
\address{\emph{${}^{a}$HUN-REN Wigner Research Centre for Physics, 1121 Budapest,
Konkoly-Thege Mikl\'os \'ut 29-33, Hungary${}^{2}$}~\\
\emph{${}^{b}$Institute of Theoretical Physics and Mark Kac Center
for Complex, Systems Research, Jagiellonian University, 30-348 Krak\'ow,
Poland}}
\begin{abstract}
We determine exact finite-volume spin form factors and two-point correlations for a massive Kramer--Wannier interface in the scaling Ising model, separating ordered and disordered phases. The interface admits two complementary lattice realizations: a self-dual inhomogeneous Ising chain whose continuum limit has a sign-changing Majorana mass profile, and, after exchanging Euclidean space and time, a mixed thermal average amenable to Toeplitz/Fredholm methods. Combining these descriptions, we obtain the exact Neveu-Schwarz-Ramond spin matrix elements and a factorized Fredholm determinant for the interface correlator. Defect TCSA and independent fermionic spin-chain calculations confirm the predicted spectrum and matrix elements.
\end{abstract}
\maketitle

\section{Introduction}

Interfaces/defects provide a non-trivial way of probing local and non-local physics in quantum many-body systems. From the non-local perspectives, defects/interfaces are known since antiquity to scatter particles, hold localized zero modes, and they are studied extensively in recent years in connection with symmetries and their effects on the large distance behavior. On the other hand, defects/interfaces also support local degrees of freedom living on them, which not only serve as 
point-like probes of large distance physics, but also allow potentially richer/subtler regularity/analytic behaviors of their correlation functions in the scaling limit. 
A particular subclass of defects/interfaces with a higher hope of exact understanding are the integrable ones, near which particles scatter purely elastically through their transmission and reflection coefficients. Such coefficients are sometimes easy to determine, but  understanding the form factors and correlators of local degrees of freedom on them are more involved than the bulk versions, and so far there are limited exact results of infinite-volume defect form factors/correlators, not to mention the finite-volume ones.

In this work, we provide the exact finite-volume spin-operator 
form factors and two-point correlators on an Ising massive interface (or domain-wall), that is closely related to the Kramers--Wannier duality transformation~\citep{KramersWannier1941}. In the continuum, it is implemented by perturbing the two sides of an ``identity defect'' in the Ising CFT by opposite thermal masses. This is a simple construction, and is {\it not} the {\it topological} duality interface associated with the KW symmetry as studied extensively in the literature~\citep{PetkovaZuber2001,FrohlichEtAl2004,AasenMongFendley2016,Shao2023}. But it appeared at least as early as in~\citep{AasenMongFendley2016}, as a way to connect the ordered and disordered regions after introducing a single duality interface in the 2D Ising model on a cylinder. Later, a Floquet version of the interface was analyzed in~\citep{Tan:2022vaz}, while an inhomogeneous-spin chain appeared in~\citep{Graf:2024itd} and contains essentially the same interface in the continuum. A fermionic action of the interface in the continuum limit was also mentioned and analyzed in~\citep{Tan:2022vaz,Seiberg:2025zqx}. However, these works focus mostly on the symmetry properties and the localized Majorana zero modes \citep{JackiwRebbi1976,Kitaev2001}, but less on the local correlators. Switching the Euclidean time and space, the same interface in the infinite-volume limit appeared in~\citep{Liu2024}, and the spin-spin two-point correlator on the interface was realized essentially as a  ``strange correlator''~\citep{You_2014} therein and was analyzed using Wiener-Hopf methods for the $2\times 2$ {\it Block} Toeplitz determinant. In fact, the diagonal elements of the Toeplitz symbol are exactly due to the Majorana zero mode and considerably complicated the analysis.  

Unfortunately, the analysis in~\citep{Liu2024} was only carried  to the un-symmetrized three-particle form-factor level and was not connected with the literature properly. In this work, we provide major improvements of the results and generalize the analysis to finite-volume. In particular, we provide a novel factorized Fredholm representation of the interface spin-spin correlator, as well the exact finite-volume form factors of the spin operator. They smoothly 
interpolate between the CFT and the infinite-volume versions. 

We further test these exact results using a defect version of the
truncated conformal space approach \citep{YurovZamolodchikov1990},
implemented with opposite thermal perturbations on the two halves
of the circle. The predicted finite-volume spectrum and the matrix
elements of the energy and spin operators are reproduced after cutoff
extrapolation. The results also confirms that the Hamiltonian spin chain and strange-correlator formulations are indeed equivalent in the continuum. Since the formulas are compact and nice, we expect the existence of simpler or deeper algebraic reasons and we hope our results could inspire further investigations.

\section{The model and interface spin correlator}

\begin{figure}
\begin{centering}
\includegraphics[width=4cm]{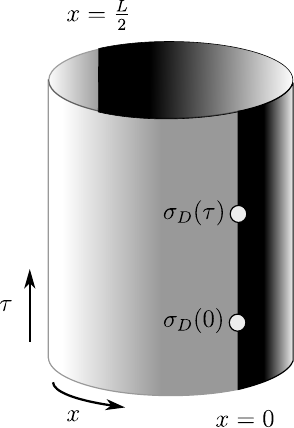}
\par\end{centering}
\caption{Interface geometry. Opposite thermal perturbations on the two halves
of the Ising cylinder produce two interfaces at $x=0$
and $x=L/2$. The defect spin fields $\sigma_{D}(0)$ and $\sigma_{D}(r)$
are inserted on the $x=0$ interface and define the spin two point correlation function. \label{Fig:cylinder}}
\end{figure}

We first introduce the continuum description of the theory on a Euclidean cylinder,
with spatial coordinate $x\sim x+L$ and Euclidean time $\tau$. The
two halves of the circle are perturbed by opposite signs of the thermal
operator, 
\[
S=S^{{\rm CFT}}_{{\rm Ising}}+\frac{m}{2\pi}\int d\tau\biggl(\int^{L/2}_{0}dx\,\varepsilon(x,\tau)-\int^{L}_{L/2}dx\,\varepsilon(x,\tau)\biggr).
\]
Equivalently, the continuum Majorana fermion has a mass profile $m(x+L/2)=-m(x)$. It can be obtained in the continuum limit of the ``closed chain'' in~\citep{Graf:2024itd}.  At the critical point the Ising admits the identity, spin-flip,
and Kramers--Wannier conformal defects \citep{OshikawaAffleck1997,PetkovaZuber2001,FrohlichEtAl2004}.
Our convention is to retain ``identity gluing'' at each interface and
encode duality in the opposite bulk masses. Thus the cylinder
contains two interfaces in an antipodal configuration, at $x=0$ and $x=L/2$, separating
regions related by Kramers--Wannier duality, see Figure \ref{Fig:cylinder}.
In what follows we measure all lengths in units of the inverse mass
and write again $L$ and $r$ for the corresponding dimensionless
circumference and separation (namely, we set $m=1$).

Our main observable is the spin-spin correlator along one of the interfaces.
We denote the spin field localized at the interface by $\sigma_{D}(\tau)\equiv\sigma(x=0,\tau)$
and study the two-point function
\[
G(r,L)=\langle\sigma_{D}(r)\sigma_{D}(0)\rangle_{L}.
\]
Since the spin operator changes the fermionic sector, its finite-volume
spectral expansion involves matrix elements between Neveu--Schwarz
and Ramond states. Using these matrix elements we write the corresponding
scaling function as 
\begin{equation}
G(r,L)=\sum_{n\in{\rm R}}\left|\langle n_{{\rm R}}|\sigma_{D}(0)|\Omega_{{\rm NS}}\rangle\right|^{2}e^{-\left(E^{{\rm R}}_{n}-E^{{\rm NS}}_{0}\right)r}.\label{eq:FFsigmasigma}
\end{equation}
 The problem is therefore to determine the finite-volume spectrum
and the interface spin form factors appearing in this expansion.

This continuum description allows two complementary realizations related to each other by switching the Euclidean time and space.   Eq.~(\ref{eq:FFsigmasigma}) treats the interface direction as temporal, this way the interface is an integrable partially transmitting Majorana defect. The corresponding lattice construction is a closed inhomogeneous Ising chain~\citep{Graf:2024itd}. Switching the directions, the interface can be realized as a mixed thermal average with the ordered
and disordered transfer matrices, generalizing the construction of~\citep{You_2014}. This leads to a block Toeplitz determinant
representation. In the continuum limit, both of the two formulations agree. 

Here we briefly summarize the two lattice constructions. For historical context on Ising correlations
and cylinder form factors, see \citep{WuEtAl1976,LeClairEtAl1996,Bugrij2001,BugrijLisovyy2003,IorgovLisovyy2011}. First, in the spin-chain formulation, the lattice Hamiltonian for a chain of length $L=2N$ reads with $0<\alpha<1$ and $X_{2N}=X_0$
\begin{align}
H=&\sum_{i=0}^{N-1}\frac{\alpha Z_i -X_iX_{i+1}}{2}+ \sum_{i=N}^{2N-1}\frac{Z_i-\alpha X_iX_{i+1}}{2} \ . \label{eq:latH}
\end{align}
This is a special case of the closed chain briefly discussed in~\citep{Graf:2024itd}. The interfaces are at $i=0$ and $N-1$. The interface spin-operator $X_0=\Gamma_0$ can be kept fermionic after fermionization. The spectrum are different in the NS and R sectors. In the Ramond sector, the fermionic Hamiltonian factorizes to two Majarona chains with un-equal odd lengths $(2N+1,2N-1)$, and the spectrum are almost the same as a homogeneous model, but with two exact Majarona zero modes, one for each chain, replacing the $\vec{p}=0$ excitation of the homogeneous chain. The excitation at the top of the spectrum also becomes ``zero-mode-like''. The one-particle Ramond excitations of the two fermion chains are labeled by $|n;\pm \rangle$, where $n\in \mathbb{Z}_{>0}$ labels the periodic quantization levels. The massive continuum limit is realized as $\alpha \rightarrow 1^-$ with all energies rescaled to $1-\alpha$. 

Equivalently, the same continuum limit can be realized through a mixed thermal average
\begin{align}
\langle X_{0}X_{n}\rangle_{L}=\lim_{M\to\infty}\frac{\mathrm{Tr}\left(e^{-\frac{L}{2}H_{\alpha^{\star}}}X_{0}X_{n}e^{-\frac{L}{2}H_{\alpha}}\right)}{\mathrm{Tr}\left(e^{-\frac{L}{2}H_{\alpha^{\star}}}e^{-\frac{L}{2}H_{\alpha}}\right)},
\end{align}
where $H_\alpha$ and $H_{\alpha^{\star}}$ are two {\it homogeneous} Ising chain Hamiltonians dual to each other under KW duality, with spatial size $M$. After Jordan--Wigner fermionization this becomes a Pfaffian, or equivalently
the square root of a $2\times2$ block analyzable Toeplitz determinant. In the
scaling limit, this determinant factorizes into Toeplitz-plus-Hankel
and Toeplitz-minus-Hankel pieces \citep{BorodinOkounkov2000,BasorWidom2000,BasorEhrhardt2002}.
This is the simplest route to the Fredholm representation below.

\section{The main results}

In this section we present the main results of the paper. The exact spin-spin correlator Eq.~(\ref{eq:FFsigmasigma}) in the continuum limit can be written as
\begin{align}
G(r,L)=S_{0}(L)^{2}e^{-\Delta(L)r}\det(1+\hat{K}_{+})\det(1+\hat{K}_{-}),
\label{eq:determin}\end{align}
where the Fredholm operators $\hat{K}_{\pm}\equiv\hat{K}_{\pm}(r,L)$
act on the finite-volume eigenstates $|n;\pm\rangle$ with $n\in \mathbb{Z}_{> 0}$. These eigenstates are labeled by their momenta, which we will see follow the free momentum quantization 
\[
p_{n}=\frac{2\pi n}{L},\qquad E_{n}=\sqrt{1+p^{2}_{n}},\qquad n=1,2,\ldots.
\]
 The corresponding kernels are
\[
\langle n|\hat{K}_{\pm}|m\rangle=\frac{2\pi}{L}\frac{k_{\pm}(p_{n})k_{\pm}(p_{m})}{\pi(E_{n}+E_{m})}e^{-\frac{r}{2}(E_{n}+E_{m})},
\]
with
\[
k_{\pm}(p)=\frac{p}{\epsilon(p)}g_{\pm}(\epsilon(p))g(\epsilon(p))^{\pm\frac{1}{2}}\quad;\quad\epsilon(p)=\sqrt{1+p^{2}} \ . 
\]
 The function $S_{0}(L)$ is the finite-volume interface one-point
function
\[
S_{0}(L)=\langle\Omega_{\mathrm{R}}|\sigma_{D}|\Omega_{\mathrm{NS}}\rangle_{L},
\]
while $\Delta(L)$ is the ground-state energy difference $\Delta(L)=E^{\mathrm{R}}_{0}(L)-E^{\mathrm{NS}}_{0}(L)$.
The appearing expressions can be very compactly expressed in terms
of 
\[
A_{L}(p)=\epsilon(p)\coth\frac{L\epsilon(p)}{2}
\]
Explicitly,
\[
\Delta(L)=\int^{\infty}_{0}\frac{dp}{2\pi}\log\left[\frac{A^{2}_{L}(p)-1}{p^{2}}\right].
\]
The defect leg functions entering the kernels are
\[
g(t)=\exp\left[\frac{2}{\pi}\int^{\infty}_{0}\frac{d\theta}{\cosh\theta}\log\frac{t+\cosh\theta}{t}\right],
\]
 and
\begin{align*}
\log g_{\pm}(t,L) & =-\int^{\infty}_{0}\frac{dp}{\pi}\frac{t}{t^{2}+p^{2}}\log\left[\frac{A_{L}(p)\mp1}{\epsilon(p)\mp1}\right] \ . 
\end{align*}
 The factors $g_{\pm}(t,L)$ tend to one as $L\to\infty$, while $g(t)$
remains as the infinite-volume defect leg factor \citep{Liu2024}.
Introduce further three even logarithms
\begin{equation}
\ell_{0}(p)=\log\frac{\epsilon(p)+1}{\epsilon(p)}\:;\quad\ell_{\pm}(p;L)=\log\frac{A_{L}(p)\pm1}{\epsilon(p)}
\end{equation}
 Then the normalized finite-volume one-point function can be written
using their Fourier transform $\tilde{f}(r)=\int^{\infty}_{-\infty}\frac{dp}{2\pi}\,e^{ipr}f(p)$
as
\begin{align}
&\log S_{0}(L)=  -\frac{1}{4}\log\sinh\frac{L}{2}+\int^{\infty}_{-\infty}\frac{dp}{4\pi}\frac{\log\frac{\epsilon(p)+1}{\sqrt{A^{2}_{L}(p)-1}}}{1+p^{2}}\label{eq:compact-E}\\
 & +\int^{\infty}_{0}\frac{rdr}{4}\left[\tilde{\ell}^{2}_{+}(r;L)+\tilde{\ell}^{2}_{-}(r;L)-2\tilde{\ell}^{2}_{0}(r)\right]+\ln S_\infty.\nonumber 
\end{align}
Several limits of these expressions are worth noting. In the ultraviolet
limit, Eq.~(\ref{eq:determin}) reduces to the determinant representation of the CFT two-point function on a cylinder~\citep{Kitanine:2011fu,Liu:2026qxw}, the ground-state splitting reproduces the CFT value $\Delta(L)=\frac{\pi}{4L}+O(L)$,  while the defect one-point function
scales as $S_{0}(L)\sim S_{\infty}C_{\sigma}L^{-1/8}.$ The exponent
is fixed by the scaling dimension of the spin field, whereas $C_{\sigma}$
is a nontrivial amplitude of the Kramers--Wannier interface~\citep{Liu2024}. In the
opposite limit $S_{0}(L)\to S_{\infty}$, while $\Delta(L)=e^{-L/2}+O(e^{-L}),$
which anticipates the localized interface mode described below. The number $S_\infty$ can be further fixed by the small-$r$ asymptotics $G(r,L) \rightarrow r^{-\frac{1}{4}}$. In
the infinite-volume limit $g_{\pm}(t,L)\to1$, and the one-particle
spin form factors reduce to
\[
F_{\pm}(\theta)\propto\tanh\theta\,g(\cosh\theta)^{\pm1/2}.
\]
Unlike the vacuum form factors usually obtained from the
integrable form-factor bootstrap that are meromorphic in $\theta$ \citep{Smirnov1992,BajnokPallaTakacs2006,BajnokElDeeb2010},
these defect-spin form factors contain a branch cut inherited from
the Wiener--Hopf factor $g$. Branch cuts are familiar in finite-temperature
and mixed-state Ising form factors \citep{Doyon2005,ChenDoyon2014},
but here they occur for a vacuum matrix element localized on a massive
interface.

Expanding the Fredholm determinants gives the exact finite-volume
spin form factors. In particular, for the two one-particle Ramond
states $|n;\pm\rangle$,
\[
|\langle n;\pm|\sigma_{D}|\Omega_{{\rm NS}}\rangle|=S_{0}(L)\frac{k_{\pm}(p_{n})}{\sqrt{LE_{n}}} \ ,
\]
 Higher form factors factorize into the corresponding product forms.
The two Fredholm determinants therefore encode two independent defect
channels, while the prefactor $e^{-\Delta(L)r}$ accounts for the
change of ground state between the NS and R sectors.  Notice that the function $g(t)$ is known in~\citep{Liu2024}, but the determinant formula is new. It agrees in the $L\rightarrow\infty$ limit with the three-particle form factor expressions in~\citep{Liu2024} after symmetrization the $dt_i$ integrals in~\citep{Liu2024}. 

\section{Integrable defect interpretation}

The interface has a continuum interpretation as an {\it integrable defect}. In the
continuum the thermally perturbed Ising model contains a free massive
Majorana fermion. The Kramers--Wannier interface corresponds to a
sign change of the mass, $m(x)=m\,{\rm sgn}(x),$ or, in finite volume,
to two antipodal sign changes. Although the bulk theory is free on
both sides, the interface is not transparent. A fermion incident on
the defect is partly reflected and partly transmitted, which is allowed
by the rapidity independent scattering \citep{DelfinoMussardoSimonetti1994}.
Imposing the continuity of the fermions at the point where the mass
changes sign gives
\[
R(\theta)=-\frac{1}{\cosh\theta},\qquad T(\theta)=-i\tanh\theta,
\]
 where $E=\cosh\theta$ and $p=\sinh\theta$. These amplitudes obey
the defect unitarity and crossing relations. In particular, low-energy particles are predominantly reflected,
while high-energy particles are transmitted. The partial transmission found here is complementary
to the perfect transmission of the topological duality interface
\citep{UedaEtAl2025}.

For the cylinder geometry with two interfaces it is useful to fold
the system along one of the defects \citep{BajnokGeorge2006}. The
problem becomes a two-component Majorana theory in an interval of
length $L/2$, with reflection matrix
\[
{\cal R}(\theta)=\left(\begin{array}{cc}
R(\theta) & T(\theta)\\
T(\theta) & R(\theta)
\end{array}\right).
\]
 Diagonalizing the two components gives two independent reflection channels
\citep{GhoshalZamolodchikov1994},
\[
R_{\pm}(\theta)=i\tanh\left(\frac{i\pi}{4}\pm\frac{\theta}{2}\right).
\]
 These are the continuum counterparts of the two channels appearing
in the Fredholm formula. The finite-volume quantization conditions
are therefore ordinary boundary Bethe--Yang equations. In the Ramond
sector the two boundaries are opposite and one finds
\[
e^{iL\sinh\theta}R_{+}(\theta)R_{-}(\theta)=1,
\]
 which reduces to the free periodic quantization except the zero modes. In the Neveu--Schwarz
sector the two channels obey
\[
e^{iL\sinh\theta}R_{\pm}(\theta)^{2}=1, 
\]
and the spectrum is non-trivial. Since the theory is free, multi-particle energies are obtained by
summing the corresponding one-particle energies.

Here we briefly comment on the zero modes. The reflection factor $R_{-}(\theta)$
has a pole at $\theta=i\pi/2$, suggesting boundary/interface bound-states localized near the two interfaces. But the nature of bound states differ in the Ramond and Neveu-Schwarz sectors. In the NS sector, the bound state is dynamical with a imaginary rapidity $\theta=iu$ determined
by
\[
e^{-L\sin u}R_{-}(iu)^{2}=1,\qquad E_{{\rm b}}=\cos u.
\]
Only for $L>2$ there is a solution with $u\in \mathbb{R}$. This mode can not appear by itself, but is responsible for the leading large-$L$ contribution
proportional to $e^{-L/2}$ in the ground-state energy difference. However, in the Ramond sector the zero mode is unconditional and always has strict zero energy ($\theta=i\pi/2$)~\citep{Seiberg:2025zqx}. 
The same picture of particle reflection/transmission can also be obtained from 
the lattice Hamiltonian Eq.~(\ref{eq:latH}). In fact, Eq.~(\ref{eq:latH}) also allows ``zero modes like'' states on top of the fermionic spectrum. For example, in the Ramond sector, the long-chain with size $2N+1$ supports an excitation with $e^{ip}=-\alpha$, whose profile is essentially the same as the low-energy zero modes at $e^{ip}=\alpha$, up to an extra $(-1)^i$. 

\section{TCSA checks}

We now compare the exact finite-volume predictions with a defect version
of the truncated conformal space approach \citep{BajnokHolloWatts2014}.
We use the Hilbert space of the critical Ising CFT on a circle and
perturb the two halves of the system with opposite signs of the energy
operator,
\[
H_{{\rm TCSA}}=H_{{\rm CFT}}+\frac{m}{2\pi}\left(\int^{L/2}_{0}dx\,\varepsilon(x)-\int^{L}_{L/2}dx\,\varepsilon(x)\right).
\]
 Equivalently, the perturbation describes a Majorana mass profile
satisfying $m(x+L/2)=-m(x)$. The two discontinuities at $x=0$ and
$x=L/2$ are the locations of the finite-volume  interfaces.

We implemented this Hamiltonian in two independent bases. In the Ising
basis the matrix elements of the perturbing field are computed from
the CFT three-point data. In the fermionic basis the same perturbation
takes a particularly transparent form: the diagonal mass terms cancel
between the two halves, while the off-diagonal momentum-changing terms
survive. The two implementations give the same truncated spectra,
providing a first internal consistency check of the defect construction.

\begin{figure}
\begin{centering}
\includegraphics[width=6cm]{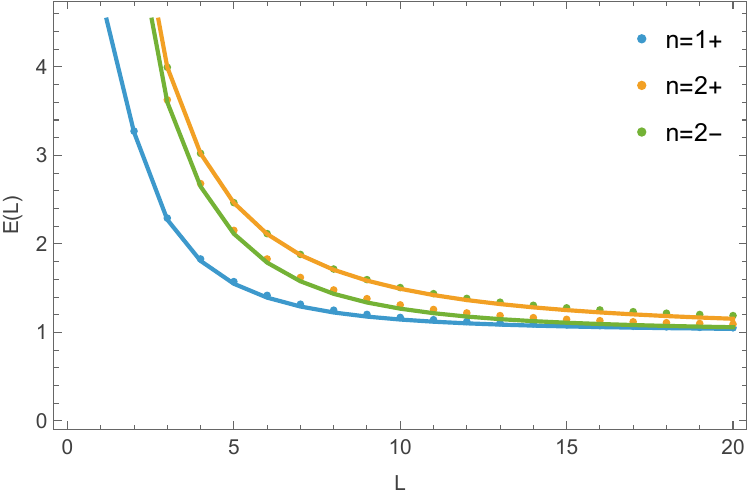}
\par\end{centering}
\caption{Low-lying even-NS spectrum. Solid lines are the exact Bethe--Yang
predictions for one-particle states in the folded defect channels,
including the localized bound-state contribution. Dots are cutoff-extrapolated
TCSA data. Labels denote the quantization number and channel, ($n,\pm$).
\label{Fig:NS}}
\end{figure}

The raw TCSA energies have visible cutoff dependence, especially in
the ground-state energy. Energy differences are considerably more
stable. We therefore compare the exact predictions with cutoff-extrapolated
quantities, treating even and odd cutoffs separately. The ground-state
splitting $\Delta(L)$ is reproduced by the extrapolated data over
the accessible range of volumes, see the Appendix.D for details. The excited-state
spectrum gives a sharper test. In the Ramond sector the one-particle
levels are doubly degenerate and follow the free periodic quantization.
In the Neveu--Schwarz sector the levels follow the two boundary-channel
quantization conditions, including the localized bound-state.
Both features are visible in the TCSA data shown in Figure \ref{Fig:NS}.

\begin{figure}
\begin{centering}
\includegraphics[width=6cm]{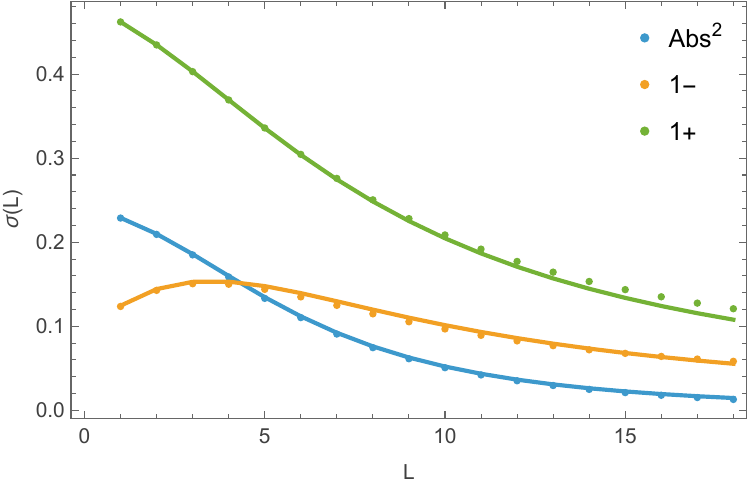}
\par\end{centering}
\caption{Defect spin form factors. Solid lines are the exact finite-volume
predictions for the one-particle matrix elements $\langle1,\pm|\sigma_{D}|\Omega_{{\rm NS}}\rangle$,
together with their squared sum. Dots are cutoff-extrapolated TCSA
data. The agreement tests the two-channel decomposition of the interface
spin response. \label{Fig:sigma}}
\end{figure}

We also tested the defect-local matrix elements by adopting techniques
in \citep{PozsgayTakacs2008,BajnokEtAl2014}. For the energy operator,
the degenerate Ramond one-particle states can be diagonalized into
the two defect channels, giving
\[
\langle n;\pm|\varepsilon_{D}|n;\pm\rangle-\langle\Omega_{\mathrm{R}}|\varepsilon_{D}|\Omega_{\mathrm{R}}\rangle=\pm\frac{2\pi}{L}\frac{p^{2}_{n}}{E^{2}_{n}},
\]
 in agreement with the numerical eigenvectors. For the spin operator,
which connects the NS and R sectors, we compared the finite-volume
matrix elements
\[
\sigma_{\pm}(L)=\langle n;\pm|\sigma_{D}|\Omega_{{\rm NS}}\rangle
\]
 with the form factors extracted from the Fredholm determinant expansion.
After normalizing by the finite-volume one-point function, the extrapolated
TCSA data agree with the predicted two-channel form-factor formulae,
see Figure \ref{Fig:sigma}. These checks confirm that the determinant
result, the continuum defect scattering picture, and the truncated
CFT calculation describe the same massive interface. We also checked numerically the finite-volume one-particle form-factors using free-fermionic methods based on Eq.~(\ref{eq:latH}), for various $1\le L(1-\alpha)\le 10$ and $167\le (1-\alpha)^{-1}\le 1000$. 

\section{Conclusion}

We have determined the exact spin responses on the ``simplest'' massive Kramers--Wannier
interface in the scaling Ising model, separating a pair of ordered/disordered regions related by the KW transformation. At on-shell level, the 
bulk theory is free, but the free-fermion transmits and reflects elastically near the interface, which also supports additional localized bound-states. The interface also allows non-Gaussian $\mathbb{Z}_2$-odd spin
operators living on them, with exactly solvable form-factors in the fermionic basis and the two point correlator. 

A central result is the factorized determinant representation of the
interface spin-spin correlator. Our simplest derivation is to use the ``strange-correlator'' type construction and utilize the special properties of the resulting block Toeplitz determinant. On the other hand, the Hamiltonian construction allows to see more detail of the finite-volume spectrum. Both constructions flow to the same defect-field-theoretical description in the continuum, serving as the basis for the truncated conformal space approach (TCSA). We have checked the exact finite-volume spectrum and defect-local matrix elements using this TCSA. The agreement
with the predicted spectrum and spin form factors provides an independent benchmark for TCSA in massive defect systems.

Our results give a solvable example of integrable interface, that is closely related to the KW transformation. It is not topological, but still symmetric enough to allow closed determinant formulas for the non-trivial interface local correlators.  
The fact that the two lattice formulations converge to the same 
field-theoretic description in the scaling limit and can be tested against TCSA also reassures a constructive eye. We hope our work 
could trigger similar
exact analyses for other duality interfaces and perturbed conformal defects, and inspire further investigations for better understanding and communication.

\acknowledgments
The authors used ChatGPT to simplify a few integrals, in particular, the integral for $S_0(L)$ in Eq.~(\ref{eq:compact-E}). The original integrals obtained without any AI usage are slightly longer and can be provided upon request.


\bibliographystyle{unsrt}
\bibliography{kw_interface_references}

\begin{widetext}

\section{Appendices}

\subsection{Continuum scattering at the identity defect with opposite masses}

We use the fermionic description of the scaling Ising field theory in the infinite volume to derive the reflection and transmission coefficients. In Minkowski signature, on each side of the defect the theory is a
free Majorana fermion,
\[
{\cal L}_{a}=\frac{i}{2}\psi_{a}(\partial_{t}+\partial_{x})\psi_{a}+\frac{i}{2}\bar{\psi}_{a}(\partial_{t}-\partial_{x})\bar{\psi}_{a}+im_{a}\bar{\psi}_{a}\psi_{a},\qquad a=L,R
\]

The two chiral components are denoted by $\psi_{a},\bar{\psi}_{a}$.
We place the defect at $x=0$, with the left theory on $x<0$ and
the right theory on $x>0$. The interface studied in the Letter is
obtained by taking
\[
m_{L}=+m,\qquad m_{R}=-m,\qquad m>0.
\]
Thus the two half-lines are related by Kramers--Wannier duality,
since the thermal perturbation changes sign under duality. 
At the conformal point, the three elementary Ising defects are labelled by
$I$, $\epsilon$, and $\sigma$.  The $I$ and $\epsilon$ defects give
transparent fermionic gluings differing by a sign, whereas the non-invertible
$\sigma$ defect implements Kramers--Wannier duality and is most naturally
described after folding
\cite{OshikawaAffleck1997,PetkovaZuber2001,FrohlichEtAl2004,AasenMongFendley2016}.
We instead impose identity gluing and implement duality through the opposite
thermal masses on the two sides.  This choice gives an unambiguous local
restriction of the bulk spin field to the interface and produces a genuinely
reflecting--transmitting massive defect.

 The bulk equations
of motion are
\[
(\partial_{t}+\partial_{x})\psi_{a}=-m_{a}\bar{\psi}_{a},\qquad(\partial_{t}-\partial_{x})\bar{\psi}_{a}=m_{a}\psi_{a}.
\]
For a uniform mass $m_{a}=\eta_{a}m$, with $\eta_{a}=\pm1$, the
one-particle solutions can be parametrized by rapidity,
\[
E(\theta)=m\cosh\theta,\qquad p(\theta)=m\sinh\theta.
\]
A convenient mode expansion is
\[
\begin{pmatrix}\begin{array}{c}
\psi_{a}(x,t)\\
\bar{\psi}_{a}(x,t)
\end{array}\end{pmatrix}=\int^{\infty}_{-\infty}\frac{d\theta}{2\pi}\left[u_{\eta_{a}}(\theta)A_{a}(\theta)e^{-iE(\theta)t+ip(\theta)x}+u^{*}_{\eta_{a}}(\theta)A^{\dagger}_{a}(\theta)e^{iE(\theta)t-ip(\theta)x}\right],
\]
where the spinors $u_{\eta}(\theta)$ are chosen to be 
\[
u_{+}(\theta)=\sqrt{m}\begin{pmatrix}\begin{array}{c}
-ie^{-\theta/2}\\
e^{\theta/2}
\end{array}\end{pmatrix},\quad u_{-}(\theta)=i\sqrt{m}\begin{pmatrix}\begin{array}{c}
ie^{-\theta/2}\\
e^{\theta/2}
\end{array}\end{pmatrix}.
\]
The relative phase in the second spinor fixes the phase convention
for the transmission amplitude below.

Consider now a particle incident from the left. For $x<0$ the field
is a superposition of the incoming and reflected waves, 
\[
\Psi_{L}(x,t)=u_{+}(\theta)e^{-iEt+ipx}+R(\theta)u_{+}(-\theta)e^{-iEt-ipx},
\]
while for $x>0$ it is a transmitted wave,
\[
\Psi_{R}(x,t)=T(\theta)u_{-}(\theta)e^{-iEt+ipx}.
\]
Here $\Psi_{a}=(\psi_{a},\bar{\psi}_{a})^{T}$. The identity defect
condition requires continuity of both fermion components at $x=0$,
$\Psi_{L}(0,t)=\Psi_{R}(0,t)$ and leads to the solution
\[
R(\theta)=-\frac{1}{\cosh\theta},\qquad T(\theta)=-i\tanh\theta.
\]
These amplitudes are a special free-fermion realization of integrable defect
scattering with simultaneous reflection and transmission
\cite{DelfinoMussardoSimonetti1994}.
The same amplitudes are obtained for incidence from the right, with
the same choice of phases. The result has a simple interpretation.
Low-energy fermions are almost completely reflected by the mass sign
change, while high-energy fermions are almost completely transmitted.
The interface is therefore neither a purely reflecting boundary nor
a topological transparent defect after the massive perturbation. Its
nontriviality comes from combining identity gluing at the conformal
point with opposite thermal perturbations on the two sides.

The amplitudes satisfy the standard integrability constraints for
a defect. Unitarity expresses conservation of probability for scattering
off the interface, including the interference between reflection and
transmission. Crossing relates the direct and crossed defect amplitudes.
In the Ising theory the bulk scattering phase is fermionic, $S=-1$,
and this fixes the relative signs in the crossing relations. These
checks are useful because the same amplitudes enter the finite-volume
quantization after folding \cite{GhoshalZamolodchikov1994,BajnokGeorge2006}.

\subsection{The block Toeplitz determinant approach}

In this section we describe the lattice construction which leads to
the determinant formula for the interface spin correlator. We use
a square-lattice regularization in which the direction along the defect
is treated as space, while the direction perpendicular to the defect
is treated as Euclidean time. The two halves of the Euclidean evolution
are governed by dual transverse-field Ising Hamiltonians 
\cite{KramersWannier1941,SchultzMattisLieb1964,Pfeuty1970}.

Let $X_{i},Y_{i},Z_{i}$ be Pauli matrices at site $i$ of a periodic
chain of length $N$, with $X_{N}=X_{0}$. We define
\[
H_{\alpha}=-\frac{1}{2}\sum^{N-1}_{i=0}X_{i}X_{i+1}+\frac{\alpha}{2}\sum^{N-1}_{i=0}Z_{i},
\]

and
\[
H_{\alpha^{\star}}=-\frac{\alpha}{2}\sum^{N-1}_{i=0}X_{i}X_{i+1}+\frac{1}{2}\sum^{N-1}_{i=0}Z_{i},
\]
where $0<\alpha<1$. The Hamiltonian $H_{\alpha}$ is in the ordered
regime, while $H_{\alpha^{\star}}$ is its Kramers--Wannier dual.
In the scaling limit $\alpha\to1$, the two Hamiltonians describe
the two signs of the Majorana mass. The corresponding infinite-volume $\alpha/\alpha^\star$ interface correlator
and its block-Toeplitz symbol were obtained in Ref.~\cite{Liu2024}; the present
construction keeps the transverse direction finite and leads to the discrete
Fredholm kernels of the Letter.

The spin response along the interface is obtained from the mixed thermal
average
\[
C_{n}(M)\equiv\langle X_{0}X_{n}\rangle_{M}=\lim_{N\to\infty}\frac{{\rm Tr}\left(e^{-\frac{M}{2}H_{\alpha^{\star}}}X_{0}X_{n}e^{-\frac{M}{2}H_{\alpha}}\right)}{{\rm Tr}\left(e^{-\frac{M}{2}H_{\alpha^{\star}}}e^{-\frac{M}{2}H_{\alpha}}\right)}.
\]
Here $n$ is the separation along the interface and $M$ is the total
Euclidean length in the transverse direction. In the continuum scaling
limit we keep
\[
r=n(1-\alpha),\qquad L=M(1-\alpha)
\]
fixed. The continuum quantity $L$ is the circumference used in the
main text. We now fermionize the spin chain. Introduce Clifford operators
\[
\Gamma_{2j}=X_{j}\prod_{0\le k<j}(-Z_{k}),\qquad\Gamma_{2j+1}=-Y_{j}\prod_{0\le k<j}(-Z_{k}),\qquad0\le j\le N-1,
\]
which satisfy $\{\Gamma_{i},\Gamma_{j}\}=2\delta_{ij}$. The spin
string becomes a string of Dirac matrices,
\[
X_{0}X_{n}=\prod^{n-1}_{j=0}\left(-i\Gamma_{2j+1}\Gamma_{2j+2}\right).
\]
Since both $H_{\alpha}$ and $H_{\alpha^{\star}}$ are quadratic in
the Majorana variables, the product $e^{-\frac{M}{2}H_{\alpha^{\star}}}e^{-\frac{M}{2}H_{\alpha}}$
defines a Gaussian fermionic density matrix. Therefore $C_{n}(M)$
is a Pfaffian of the two-point contractions of the Majoranas
\cite{Kaufman1949,SchultzMattisLieb1964,WuEtAl1976,Liu2024}. Let
$\eta_{j}=\begin{pmatrix}\Gamma_{2j+1},\Gamma_{2j+2}\end{pmatrix}$
with $j=0,\ldots,n-1$. The relevant contraction matrix has block
entries depending only on $i-j$ in the thermodynamic limit. Thus
the square of the Pfaffian is a block Toeplitz determinant, $C_{n}(M)^{2}=\det T_{n}(a),$
where
\[
T_{n}(a)=\left([a]_{i-j}\right)_{0\le i,j\le n-1}
\]
is an $n\times n$ block Toeplitz matrix with $2\times2$ matrix entries.
The Fourier coefficients are
\[
[a]_{k}=\oint_{|z|=1}\frac{dz}{2\pi iz}\,z^{k}a(z)
\]
The sign of $C_{n}(M)$ is fixed by continuity from $C_{0}(M)=1$.
The block symbol is
\[
a(z)=\frac{1}{1-\frac{16wz(1-\alpha z)(1-\alpha z^{-1})}{(1-w)^{2}(1-z)^{2}(1+\alpha)^{2}}}\begin{pmatrix}\begin{array}{cc}
A(z) & B(z)\\
zB(z) & A(z)
\end{array}\end{pmatrix},
\]
where
\[
A(z)=\frac{1-\alpha}{1+\alpha}\frac{1+z}{1-z}
\]
\[
B(z)=\frac{2\sqrt{(1-\alpha z)(1-\alpha z^{-1})}}{(1+\alpha)(1-z)}\frac{1+w}{1-w},
\]
and
\[
w=e^{-M\epsilon(z)},\qquad\epsilon(z)=\sqrt{(1-\alpha z)(1-\alpha z^{-1})}.
\]
The branch of the square root is chosen such that $\epsilon(1)=1-\alpha>0$. A useful relation is

\[
\det a(z) \equiv f_0(z)=\left[1-\frac{16wz(1-\alpha z)(1-\alpha z^{-1})}{(1-w)^{2}(1-z)^{2}(1+\alpha)^{2}}\right]^{-1}.
\]
For $M\to\infty$, one has $w\to0$, and the symbol reduces to the
infinite-volume $\alpha/\alpha^{\star}$ version. A crucial feature of the kernel is that it exponentiates through a {\it first order} polynomial matrix
\[
J(z)=\left(\begin{array}{cc}
0 & 1\\
z & 0
\end{array}\right)
\]
as
\[
a(z)=\sqrt{f_{0}(z)}\,\exp\{f(z)J(z)\}.
\]
where $f(z)$ and $f_0(z)$ are scalar functions. This implies that for windingless $f(z)$ and $f_0(z)$, $a(z)$ allows commutative Wiener--Hopf factorization through the additive versions of the scalar functions, which is indeed the case in the infinite volume after shifting the contour to $|z|=1^-$. 

At finite-volume, the properties of $f_0(z)$ and $f(z)$ becomes more complicated near $|z|=1$. For $f_0(z)$, there is a double-pole at $z=1$, while $f(z)$ has a jump at $z=1$
\begin{align}
f(e^{i0^+})-f(e^{i0^-})=\pi i  \ . \nonumber 
\end{align}
This  prevents a direct use of the Wiener--Hopf method. To obtain the scaling Fredholm formula, we can use the following approximation which does not change all the functions in the scaling region $1-z={\cal O}(1-\alpha)$
\begin{align}
&J(z) \rightarrow \sigma_1 \ , \nonumber \\ 
& f_0(z)^{-1} \rightarrow 1-\frac{4w z (1-\alpha z)(1-\alpha \bar z)}{(1-w)^2(1-z)^2} \ , \nonumber \\ 
& f(z) \rightarrow \ln \bigg(1+\frac{\sqrt{1-\alpha z} \sqrt{1-\alpha \bar z}}{1-\alpha}\frac{1+w}{1-w}\bigg)-\ln \frac{1+\alpha}{1-\alpha}-\ln \frac{1-z}{1+z}+\frac{1}{2}\ln f_0 \ . \nonumber 
\end{align}
This simplification is not equivalent to the original block determinant at generic $\alpha$, but preserves the anti-symmetry property of the original matrix. Moreover,  when one circles around $|z|=1^-$, the jumps of $f_0(z)$ and $f(z)$ are preserved in this approximation, providing a strong justification of the approximation. After this approximation, since $\sigma_1$ is a constant, we can chose the diagonal basis, leading to two factorized scalar determinants with kernels
\begin{align}
&c_1(z)=\frac{1+z}{1-z} f_0(z)  \left(1-\alpha+\frac{1+w}{1-w} \epsilon(z)\right)   \ , \nonumber \\ 
&c_2(z)=\frac{1-z}{1+z} \left(1-\alpha+\frac{1+w}{1-w} \epsilon(z)\right)^{-1} \ . \nonumber 
\end{align}
As such, in the scaling region one expects
\begin{align}
\langle X_0X_{2n}\rangle^2 \rightarrow \det T_{2n}(c_1) \det T_{2n}(c_2) \ . \nonumber 
\end{align}
Here we chose the separation number $2n$ to be even, since in this case the contour with ${\rm PV}$ at $|z|=1$ for the approximated kernels can be deformed to $|z|=1^-$ which share the same winding numbers as the original kernel, without changing the value of the determinants. The final scaling function is independent of
this auxiliary choice. To further simplify, we notice that  one can write
\begin{align}
&c_1(z)=\frac{1+z}{1-z}c(z) \ ,  \nonumber \  
c_2(z)=\frac{1-z}{1+z}  d(z) \ ,  \nonumber \\ 
&c(z)=c(z^{-1}) \ , \ d(z)=d(z^{-1}) \ . \nonumber 
\end{align}
For the kernels of the form above, there are two crucial algebraic relations \cite{BasorEhrhardt2000,BasorEhrhardt2002} which relate the Toeplitz determinants to Toeplitz $\pm$ Hankel determinants for symmetric kernels. 
\begin{align}
&(\det T_{2n}(c_1))=\det (T_n( c)+H_n( c)) ^2\ , \nonumber \\
&(\det T_{2n}(c_2))=\det (T_n(d)-H_n( d)) ^2\ ,\nonumber 
\end{align}
where
\[
H_{n}(c)=\left([c]_{i+j+1}\right)_{0\le i,j\le n-1} \ ,
\]
is the Hankel determinant. The two identities relate to each other through the transformation $c(z) \rightarrow c(-z)$ that changes $c_{i-j} \rightarrow (-1)^{i-j}c_{i-j}$. 
These are the steps which explain the factorization into two determinant
channels. There is one more technical point. The symbol $c(z)$ has a winding number associated with the singularity at $z=1$. To apply the
Wiener--Hopf/BOCG formulae \cite{BorodinOkounkov2000,BasorWidom2000} one first regularizes
\[
(1-z)(1-z^{-1})\quad\longrightarrow\quad(1-\beta z)(1-\beta z^{-1}),\qquad0<\beta<1,
\]
performs the factorization at fixed $\beta$, and takes $\beta\to1$
at the end \cite{BasorEhrhardt1998,BasorEhrhardt2005}. 
At finite $\beta$, the kernel $c(z)$ along the unite circle becomes zero-winding, and one can use the standard identities in the theory of Toeplitz--Hankel
and Wiener--Hopf--Hankel determinants
\cite{BasorEhrhardt1998,BasorEhrhardt2000,BasorEhrhardt2005}; here we retain
only the part required to obtain the finite scaling kernel. After Wiener--Hopf
factorization and contour deformation, the Fredholm kernels are obtained
from the poles
\[
e^{-L\sqrt{1+p^{2}}}=1,\qquad p=iE_{n},\qquad E_{n}=\sqrt{1+p^{2}_{n}},\qquad p_{n}=\frac{2\pi n}{L},\qquad n=1,2,\ldots.
\]
These leads to the final finite-volume determinant representation
in the main text. 

\subsection{The inhomogeneous spin-chain}
In this section, we consider the second approach, in which the direction along
the interface is treated as Euclidean time 
\cite{SchultzMattisLieb1964,Pfeuty1970,AasenMongFendley2016}. This gives a direct Hamiltonian
interpretation of the finite-volume spectrum and of the spin form
factors. The lattice version has the following Hamiltonian
\begin{align}
H=&\frac{1}{2}\sum_{i=0}^{N-1}\left(\alpha Z_i -X_iX_{i+1}\right)+ \frac{1}{2}\sum_{i=N}^{2N-1}\left(Z_i-\alpha X_iX_{i+1} \right)  \ . \nonumber 
\end{align}
This is a special case of the closed chain discussed in~\citep{Graf:2024itd}.  We summarize its diagonalization and the resulting spectrum, since it also identifies the two defect channels appearing in the determinant formula.

We again using the standard fermionization, with  Majorana variables $\Gamma_{i}$, with $\{\Gamma_{i},\Gamma_{j}\}=2\delta_{ij}$.
The defect spin is $\sigma_{D}=X_0=\Gamma_{0}$. The two halves of the chain
are labeled by positive and negative indices. For a system with $2N$
lattice spacings between the two interfaces, the quadratic Hamiltonians
in the two charge sectors are
\begin{align*}
2H_{\pm} & =i\alpha\Gamma_{0}\Gamma_{1}+i\alpha\sum^{N-1}_{k=1}\Gamma_{2k}\Gamma_{2k+1}+i\sum^{N-1}_{k=1}\Gamma_{2k-1}\Gamma_{2k}+i\Gamma_{2N-1}\Gamma_{2N}+i\Gamma_{2N}\Gamma_{-(2N-1)}\\
 & \hspace{1.4cm}+i\alpha\sum^{N-1}_{k=1}\Gamma_{-(2k+1)}\Gamma_{-2k}+i\sum^{N-1}_{k=1}\Gamma_{-2k}\Gamma_{-(2k-1)}\mp i\alpha\Gamma_{-1}\Gamma_{0}.
\end{align*}
Here $\alpha<1$, and the scaling limit is $\alpha\to1$. The Clifford variables on the left are relabeled to negative values. 

The full Hamiltonian
is obtained by projecting with the $\mathbb{Z}_{2}$ charge,
\[
H=H_{+}\frac{1+Z}{2}+H_{-}\frac{1-Z}{2}.
\]
We identify the charge-even sector with the Neveu--Schwarz sector
and the charge-odd sector with the Ramond sector. To proceed further we chose $N=1\mod 2$, but the other case is similar. In this case,  we make the following redefinition
\[
\widetilde{\Gamma}_{i}=(-1)^{i-1}\Gamma_{-i},\qquad1\le i\le2N-1,
\]
and then taking symmetric and antisymmetric combinations
\[
\Gamma_{a,i}=\frac{\Gamma_{i}+\widetilde{\Gamma}_{i}}{\sqrt{2}},\qquad\Gamma_{b,i}=\frac{\Gamma_{i}-\widetilde{\Gamma}_{i}}{\sqrt{2}}.
\]
In this basis the Hamiltonians in the two sectors all factorize into two independent Majorana
chains. In the NS sector, $\Gamma_{0}$ couples to the $a$-chain, while
the far endpoint $\Gamma_{2N}$ couples to the $b$-chain. In the R
sector the role of $\Gamma_{0}$ is exchanged: it couples to the $b$-chain, making the two chains of un-equal odd lengths. This two chains are the lattice origin of the two defect channels. 

In each of the chains in each sector, the spectrum can be found by diagonalizing the skew-symmetric rotation matrices for the Clifford bilinears. The diagonalization can be performed using the plane-wave eigenvector ansatz with reflection terms
\[
v_{2k}=\alpha_{q}e^{iqk}+R_{q}\alpha_{-q}e^{-iqk},\qquad v_{2k+1}=\beta_{q}e^{iqk}+R_{q}\beta_{-q}e^{-iqk} . 
\]
The bulk equation solves $E_q$, $\alpha_q$ and $\beta_q$, while the equations at the endpoints lead to the reflection coefficient $R_{-q}$ and the quantization condition. In the NS sector one obtains two quantization conditions,
\[
z^{2N}=\frac{(1-hz^{-1})(1+hz)}{(1+hz^{-1})(1-hz)}\qquad(a\text{-chain}),
\]
and
\[
z^{2N}=\frac{(1+hz^{-1})(1-hz)}{(1-hz^{-1})(1+hz)}\qquad(b\text{-chain}),
\]
where $z=e^{iq}$. In the continuum scaling limit, $q=(1-h)k$ and
$L=2N(1-h)$, these become
\[
e^{ikL}=\frac{i-k}{i+k},\qquad e^{ikL}=\frac{i+k}{i-k}.
\]
These are precisely the two boundary-channel quantization conditions
obtained from the folded continuum defect 
\cite{GhoshalZamolodchikov1994,BajnokGeorge2006}. The first equation also
contains the localized low-energy solution, which becomes the bound-state
branch in the continuum description. In the R sector the propagating
modes satisfy the free periodic condition $z^{2N}=1,$ up to the zero-mode
solutions. Thus the R-sector one-particle
spectrum is doubly degenerate in the scaling limit,
\[
E_{n}=\sqrt{1+\left(\frac{2\pi n}{L}\right)^{2}},\qquad n=1,2,\ldots.
\]
This is the same spectrum used in the finite-volume form-factor expansion. Here we comment on the zero modes. Since the two chains are of odd sizes, each chain's skew-symmetric rotation matrix supports a null-vector. In the short chain $a$ with total size $2N-1$, the null-vector is supported near $\Gamma_{2N-1}$
\begin{align}
h\Gamma_3=\Gamma_1 \ , h\Gamma_5=\Gamma_3,.... \ ,
\Gamma_{2k+1}=\Gamma_{2N-1}\left(h\right)^{N-1-k}, \ k=1,..N-1 \ . \nonumber  
\end{align}
In the longer chain $b$ with size $2N+1$, the null-vector is supported near $\Gamma_0$
\begin{align}
\Gamma_2=\sqrt{2}h \Gamma_0, \Gamma_4=h\Gamma_2,... \ , \Gamma_{2N}=\frac{h}{\sqrt{2}}\Gamma_{2N-2} \ . \nonumber 
\end{align}
The two null vectors combine to produce a zero-mode excitation $\{c_0,c_0^\dagger \}=1$ with an appropriate linear combination determined by the charge condition.  In addition, the chain $b$ also supports a ``high-energy zero mode'' at $z=-\alpha$ as a $2\times 2$ block in the canonical-form of the rotation matrix over reals. The eigenvector is a combination of the zero-mode profiles given above, with extra $(-1)^j$ factors.

The defect spin operator $\sigma_{D}=\Gamma_{0}$ changes the charge
sector and therefore connects NS and R states. If $|\Omega_{{\rm NS}}\rangle$
and $|\Omega_{{\rm R}}\rangle$ denote the two ground states, the
one-particle form factors are of the form
\[
f_{a}(q)=\langle\Omega_{{\rm R}}|c_0c_{a}(q)\Gamma_{0}|\Omega_{{\rm NS}}\rangle,\qquad f_{b}(q)=\langle\Omega_{{\rm R}}|c_0c_{b}(q)\Gamma_{0}|\Omega_{{\rm NS}}\rangle.
\]
The two amplitudes $f_{a}$ and $f_{b}$ are the lattice representatives
of the two continuum channels, denoted by $+$ and $-$ in the main text.
This formulation allows to check the analytic results in the scaling limit using free-fermionic numerical method (which requires inverting $4N\times 4N$ conversion matrices between NS and R sectors). The results agree perfectly.  Finally, the ground-state energy difference \cite{Bajnok:2004tq} is obtained as the difference
of zero-point energies in the two charge sectors, $\Delta(L)=E_{{\rm R}}(L)-E_{{\rm NS}}(L).$
After taking the scaling limit, this agrees with the determinant result
quoted in the main text. Thus the self-dual chain gives a direct spectral
interpretation of the same finite-volume spin response derived from
the mixed thermal average.

\subsection{Truncated conformal space checks}

In this section we describe the defect truncated conformal space approach
used to check the exact finite-volume predictions. The calculation
is performed in the Hilbert space of the critical Ising CFT on a circle.
The perturbation has opposite signs on the two halves of the circle,
producing two antipodal interfaces at $x=0$ and $x=L/2$. We use
the angular coordinate 
\begin{equation}
\theta=\frac{2\pi x}{L},\qquad0\leq\theta<2\pi,\tag{S1}
\end{equation}
and map the cylinder to the plane by 
\begin{equation}
z=e^{\frac{2\pi}{L}(\tau+ix)},\qquad\bar{z}=e^{\frac{2\pi}{L}(\tau-ix)}.\tag{S2}
\end{equation}
At $\tau=0$, this gives $z=e^{i\theta}$, $\bar{z}=e^{-i\theta}$.
The dimensionless TCSA Hamiltonian is 
\begin{equation}
h_{{\rm TCSA}}\equiv\frac{L}{2\pi}H_{{\rm TCSA}}=L_{0}+\bar{L}_{0}-\frac{c}{12}+\frac{mL}{2\pi\kappa}\left[\int^{\pi}_{0}d\theta\,\varepsilon(e^{i\theta},e^{-i\theta})-\int^{2\pi}_{\pi}d\theta\,\varepsilon(e^{i\theta},e^{-i\theta})\right].\tag{S3}
\end{equation}
Here $c=1/2$. The constant $\kappa$ fixes the normalization between
the CFT energy field and the fermion bilinear; in the fermionic convention
used below, $\kappa=2\pi$. All numerical comparisons are made in
units $m=1$.

The Ising Hilbert space is decomposed into Neveu--Schwarz and Ramond
sectors 
\cite{DiFrancescoMathieuSenechal1997}, 
\begin{equation}
{\cal H}_{{\rm NS}}={\cal V}_{0}\otimes\bar{{\cal V}}_{0}\oplus{\cal V}_{1/2}\otimes\bar{{\cal V}}_{1/2},\qquad{\cal H}_{{\rm R}}={\cal V}_{1/16}\otimes\bar{{\cal V}}_{1/16}.\tag{S4}
\end{equation}
The energy perturbation preserves the sector, while the spin operator
connects the two sectors. In practice we truncate the basis by 
\begin{equation}
L_{0}+\bar{L}_{0}\leq E_{{\rm cut}}.\tag{S5}
\end{equation}
If a non-orthonormal Virasoro basis is used, the Hamiltonian is diagonalized
as a generalized eigenvalue problem with the CFT Gram matrix. Let
$|i\rangle,|j\rangle$ be CFT states of momenta $P_{i},P_{j}$, and
define 
\begin{equation}
\varepsilon_{ij}=\langle i|\varepsilon(1,1)|j\rangle.\tag{S6}
\end{equation}
The coordinate dependence of a primary field gives 
\begin{equation}
\langle i|\varepsilon(e^{i\theta},e^{-i\theta})|j\rangle=e^{i\theta(P_{j}-P_{i})}\varepsilon_{ij}.\tag{S7}
\end{equation}
Writing $k=P_{j}-P_{i}$, the integral over the first half-circle
is 
\begin{equation}
\int^{\pi}_{0}d\theta\,\langle i|\varepsilon(e^{i\theta},e^{-i\theta})|j\rangle=\begin{cases}
\dfrac{(-1)^{k}-1}{ik}\,\varepsilon_{ij}, & k\neq0,\\[1.1em]
\pi\,\varepsilon_{ij}, & k=0.
\end{cases}\tag{S8}
\end{equation}
The second half-circle gives the opposite off-diagonal contribution
and the same diagonal contribution. Therefore the opposite-sign perturbation
in (S3) cancels the diagonal matrix elements and doubles the off-diagonal
ones: 
\begin{equation}
V_{ij}=\frac{mL}{2\pi\kappa}\begin{cases}
2\dfrac{(-1)^{k}-1}{ik}\,\varepsilon_{ij}, & k\neq0,\\[1.1em]
0, & k=0.
\end{cases}\tag{S10}
\end{equation}
This is the matrix used in the Ising-basis implementation. 

As an independent check we also implemented the same Hamiltonian directly
in the fermionic basis. The massless Majorana fields on the circle
are expanded as 
\begin{equation}
\psi(x,t)=\sqrt{\frac{2\pi}{L}}\sum_{n\in{\mathbb{Z}}+\nu}b_{n}e^{\frac{2\pi in}{L}(x-t)},\qquad\bar{\psi}(x,t)=\sqrt{\frac{2\pi}{L}}\sum_{n\in{\mathbb{Z}}+\nu}\bar{b}_{n}e^{-\frac{2\pi in}{L}(x+t)},\tag{S11}
\end{equation}
where $\nu=1/2$ in the NS sector and $\nu=0$ in the R sector. With
the convention 
\begin{equation}
\varepsilon=\frac{i}{\kappa}\psi\bar{\psi},\qquad\kappa=2\pi,\tag{S12}
\end{equation}
a uniform thermal perturbation gives 
\begin{equation}
i\int^{L}_{0}dx\,\psi(x,0)\bar{\psi}(x,0)=2\pi i\sum_{n}b_{n}\bar{b}_{n}.\tag{S13}
\end{equation}
For the half-circle one obtains 
\begin{equation}
i\int^{L/2}_{0}dx\,\psi(x,0)\bar{\psi}(x,0)=\sum_{n\neq m}b_{n}\bar{b}_{m}\frac{(-1)^{n-m}-1}{n-m}+i\pi\sum_{n}b_{n}\bar{b}_{n}.\tag{S14}
\end{equation}
The contribution from the second half-circle has the opposite off-diagonal
part and the same diagonal part. Thus the opposite-sign perturbation
removes the diagonal mass term and leaves only the off-diagonal mixing,
\begin{equation}
h^{{\rm ferm}}_{{\rm TCSA}}=L_{0}+\bar{L}_{0}-\frac{c}{12}+\frac{mL}{2\pi\kappa}\,2\sum_{n\neq m}b_{n}\bar{b}_{m}\frac{(-1)^{n-m}-1}{n-m}.\tag{S15}
\end{equation}
The spectra obtained from (S10) and (S15) agree within numerical precision.
This provides a useful check on the CFT matrix elements and on the
treatment of the two fermionic sectors.

The cutoff dependence is visible in the absolute ground-state energies.
We therefore compare primarily cutoff-extrapolated energy differences
and normalized matrix elements. For each volume we fit even and odd
cutoffs separately \cite{GiokasWatts2011}, using 
\begin{equation}
Q(E_{{\rm cut}})=Q_{\infty}+\frac{a_{1}}{E_{{\rm cut}}}+\cdots.\tag{S16}
\end{equation}
In the range used for the main-text comparison, the linear fit in
$1/E_{{\rm cut}}$ is sufficient to stabilize the low-lying energy
differences. The residual spread between the even and odd extrapolations
is used as a simple estimate of the truncation uncertainty.

The first spectral check is the ground-state splitting 
\begin{equation}
\Delta(L)=E_{{\rm R}}(L)-E_{{\rm NS}}(L).\tag{S17}
\end{equation}
The extrapolated TCSA data are compared with the exact expression
quoted in the main text on Figure \ref{Fig:E0diff}. The agreement
is good for the range of volumes shown there. At larger volumes the
extrapolation becomes less stable, as expected for a relevant bulk
perturbation in TCSA.

\begin{figure}
\begin{centering}
\includegraphics[width=6cm]{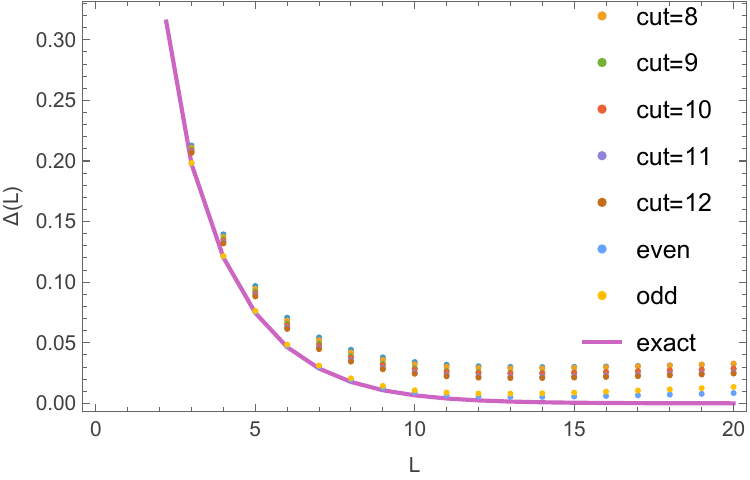}
\par\end{centering}
\caption{TCSA check of the ground-state splitting $\Delta(L)=E^{{\rm R}}_{0}(L)-E^{{\rm NS}}_{0}(L)$. Colored
points show raw TCSA data at several truncation cutoffs. The even-
and odd-cutoff sequences are extrapolated separately, giving the two
extrapolated data sets shown. The solid line is the exact finite-volume
result. The agreement confirms the predicted NS--R ground-state
energy difference entering the exponential prefactor of the interface
spin correlator. \label{Fig:E0diff}}
\end{figure}

The excited-state spectra give a more detailed test. In the Ramond
sector the one-particle energies are doubly degenerate and obey the
free periodic quantization
\begin{equation}
E_{n}=\sqrt{1+\left(\frac{2\pi n}{L}\right)^{2}},\qquad n=1,2,\ldots.\tag{S18}
\end{equation}
Many-particle energies are sums of these one-particle energies, see
Figure \ref{Fig:R}. 

\begin{figure}
\begin{centering}
\includegraphics[width=6cm]{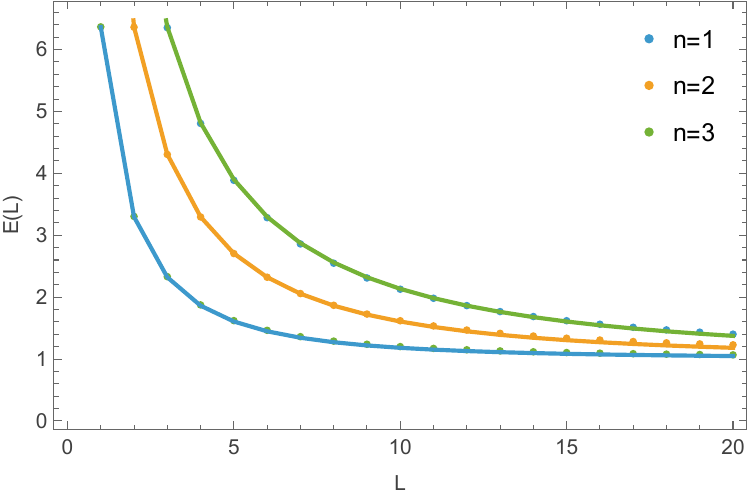}
\par\end{centering}
\caption{Low-lying R spectrum. Solid lines are the exact Bethe-{}-Yang predictions
for one-particle states in the folded defect channels. Dots are cutoff-extrapolated
TCSA data. Labels denote the quantization number. \label{Fig:R}}
\end{figure}

In the Neveu--Schwarz sector the two one-particle branches obey the
folded-channel quantization conditions
\begin{equation}
e^{iL\sinh\theta}R_{+}(\theta)^{2}=1,\qquad e^{iL\sinh\theta}R_{-}(\theta)^{2}=1,\tag{S19}
\end{equation}
with
\begin{equation}
R_{\pm}(\theta)=i\tanh\left(\frac{i\pi}{4}\pm\frac{\theta}{2}\right).\tag{S20}
\end{equation}
The $R_{-}$ channel also contains the localized branch with imaginary
rapidity $\theta=iu$,
\begin{equation}
e^{-L\sin u}R_{-}(iu)^{2}=1,\qquad E_{{\rm b}}=\cos u.\tag{S21}
\end{equation}
The NS TCSA spectrum reproduces both the scattering branches and this
localized state see Figure \ref{Fig:NSodd}.

\begin{figure}
\begin{centering}
\includegraphics[width=6cm]{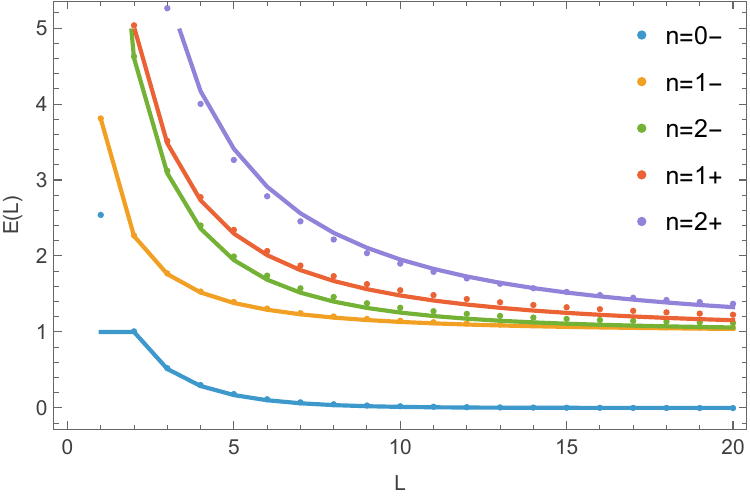}
\par\end{centering}
\caption{Low-lying odd-NS spectrum. Solid lines are the exact Bethe-{}-Yang
predictions for one-particle states in the folded defect channels,
including the localized bound-state contribution. Dots are cutoff-extrapolated
TCSA data. Labels denote the quantization number and channel, ($n,\pm$).
\label{Fig:NSodd}}
\end{figure}

We also computed defect-local matrix elements. The local fields on
the cylinder are related to plane fields by 
\begin{equation}
\varepsilon_{{\rm cyl}}(0,0)=\left(\frac{2\pi}{L}\right)\varepsilon_{{\rm plane}}(1,1),\qquad\sigma_{{\rm cyl}}(0,0)=\left(\frac{2\pi}{L}\right)^{1/8}\sigma_{{\rm plane}}(1,1).\tag{S22}
\end{equation}
The energy field preserves the fermionic sector, while the spin field
changes it.

In the Ramond one-particle subspace, the two states are degenerate
before choosing a defect-channel basis. We fix this basis by diagonalizing
the matrix of the defect energy operator. After subtracting the Ramond
vacuum contribution, the predicted diagonal matrix elements are 
\begin{equation}
\langle n;\pm|\varepsilon_{D}|n;\pm\rangle_{{\rm R}}-\langle\Omega_{{\rm R}}|\varepsilon_{D}|\Omega_{{\rm R}}\rangle=\pm\frac{2\pi}{L}\tanh^{2}\theta_{n},\tag{S23}
\end{equation}
where 
\begin{equation}
\sinh\theta_{n}=\frac{2\pi n}{L}.\tag{S24}
\end{equation}
This comparison fixes the relative identification of the $+$ and
$-$ one-particle states. 

The form factors obtained from TCSA are compared to the analytical values on Figure \ref{Fig:eps}.

\begin{figure}
\begin{centering}
\includegraphics[width=6cm]{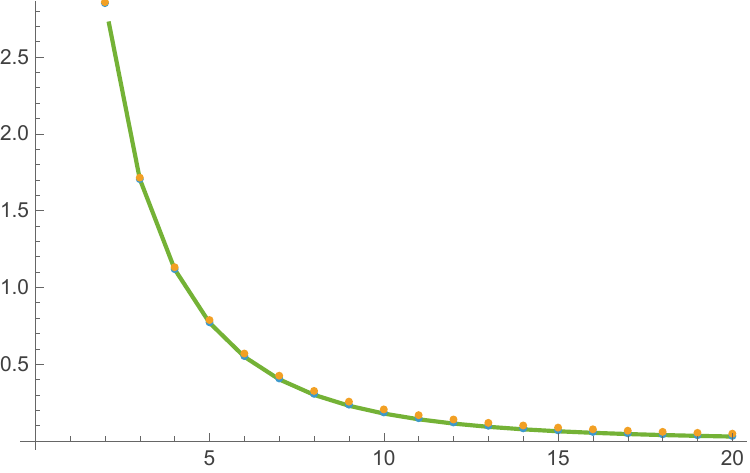}
\par\end{centering}
\caption{The form factor combination appearing in eq. (S23) as measured in TCSA against the volume with dots and its analytical form with a solid line. 
\label{Fig:eps}}
\end{figure}

Finally, we compare the spin form factors. Since $\sigma_{D}$ connects
the NS and R sectors, we compute 
\begin{equation}
\langle n;\pm|\sigma_{D}|\Omega_{{\rm NS}}\rangle.\tag{S25}
\end{equation}
We start by checking the finite-volume
one-point matrix element \begin{equation}
s_{0}(L)=\langle\Omega_{{\rm R}}|\sigma_{D}|\Omega_{{\rm NS}}\rangle.\tag{S26}
\end{equation}
The results for the cutoff extrapolated TCSA data are presented on Figure \ref{Fig:sigma11theory} with an excellent agreement.

\begin{figure}
\begin{centering}
\includegraphics[width=6cm]{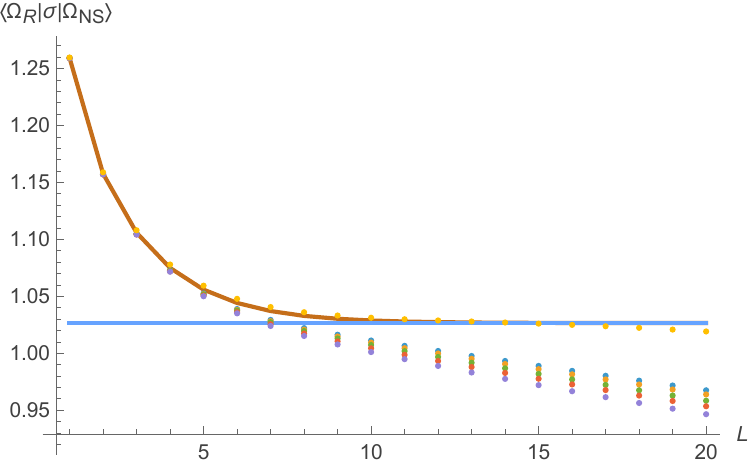}
\par\end{centering}
\caption{The finite volume one-point function of $\sigma$ against the volume. Dots represent the TCSA data and their cutoff extrapolation. Solid lines are the theoretical finite volume  expectation value and its infinite volume asymptotics.  
\label{Fig:sigma11theory}}
\end{figure}
To reduce cutoff and normalization effects, we divide the general finite volume form factors by $s_0(L)$. 
The quantities compared to the exact determinant prediction are therefore
\begin{equation}
{\cal F}^{\pm}_{n}(L)=\left|\frac{\langle n;\pm|\sigma_{D}|\Omega_{{\rm NS}}\rangle}{\langle\Omega_{{\rm R}}|\sigma_{D}|\Omega_{{\rm NS}}\rangle}\right|^{2}.\tag{S27}
\end{equation}
The exact formula gives 
\begin{equation}
{\cal F}^{+}_{n}(L)=\frac{2\pi}{L}\frac{1}{\pi}\frac{p^{2}_{n}}{E^{2}_{n}}\frac{\widetilde{g}(E_{n})g^{2}_{+}(E_{n},L)}{2E_{n}},\tag{S28}
\end{equation}
and 
\begin{equation}
{\cal F}^{-}_{n}(L)=\frac{2\pi}{L}\frac{1}{\pi}\frac{p^{2}_{n}}{E^{2}_{n}}\frac{g^{2}_{-}(E_{n},L)}{\widetilde{g}(E_{n})\,2E_{n}},\tag{S29}
\end{equation}
with 
\begin{equation}
p_{n}=\frac{2\pi n}{L},\qquad E_{n}=\sqrt{1+p^{2}_{n}}.\tag{S30}
\end{equation}
Within the cutoff extrapolation, the TCSA matrix elements agree with
(S28), (S29). This confirms not only the finite-volume spectrum, but
also the two-channel form-factor content of the determinant representation.

\end{widetext}

\end{document}